\documentclass[
    ,final            
    ,sort&compress    
    ,numberedheadings 
  ]
  {aipproc}

\usepackage{hyperref}           
\usepackage[latin1]{inputenc}   

\layoutstyle{8x11single}

\providecommand{\email}[1]{E-mail:
\href{mailto:#1}{\textup{\texttt{#1}}}}

\begin{document}

\title{Was There a Decelerating Past for the Universe?}


\author{Moncy V. John}{
  address={Department of Physics, St. Thomas College, Kozhencherri - 689641, Kerala State, India., \email{moncy@iucaa.ernet.in}}
}



\begin{abstract}
Some analyzes of the apparent magnitude-redshift data of type Ia supernovas indicate that the suspected dark energy in the universe cannot  be regarded as a  cosmological constant of general relativistic origin or as the vacuum energy encountered in quantum field theories. If this is the case, our knowledge of the physical world  remains deficient since no tested theory involves such a dark energy. Under this circumstance, an equation of state of the form $p=w\rho$ is not well-motivated and one is unable to use the Einstein equation in this case as well.   I argue that the very method of analysing the data by assuming exotic energy densities with strange equations of state itself is  misleading and the  reasonable remaining option  is to make a model-independent analysis of SNe data, without reference to the energy densities. In this basically kinematic approach, we limit ourselves to the observationally justifiable assumptions of homogeneity and isotropy, i.e., to the assumption that the universe has a RW metric. This cosmographic approach is historically the original one in cosmology. The analysis was performed  by expanding the scale factor into a fifth-order polynomial, an assumption that can be further generalized to any order. The values obtained for the present expansion rates $h$, $q_0$, $r_0$ etc. are  relevant, since any cosmological solution would ultimately need to explain them. 

Using this method, we   address an important question relevant to cosmology: Was there a decelerating past for the universe? To answer this,   the Bayes's probability  theory is employed, which is the most appropriate tool for quantifying our knowledge when it changes through the acquisition of new data. The cosmographic approach helps  to sort out   models which were always accelerating  from those which  decelerated for at least some time in the period of interest.  Bayesian model comparison technique is used to discriminate these rival hypotheses with the aid of recent releases of supernova data.  It is argued that the lessons learned using Bayesian theory are extremely valuable to avoid frequent U-turns in cosmology.

\end{abstract}

\maketitle


\section{Introduction}

\label{sec:1}

Modern cosmology began with the Hubble's discovery that there is a systematic relationship between the apparent magnitudes and redshifts of galaxies. He observed that as galaxies become fainter and fainter, their redshifts increase systematically. When these redshifts are interpreted as originating from a Doppler-like effect, one concludes that farther galaxies recede from us with greater speeds. Assuming that we do not enjoy any preferred position in the universe, this discovery naturally leads to the concept of an expanding universe. Ever since 1929, the expansion is the primary piece of information in cosmology. In this talk, my starting premises are that of an expanding, homogeneous and isotropic  universe.

The Friedman solution of Einstein equation in cosmology, which assumes that at large scales the universe has a Robertson-Walker metric and is filled with ordinary relativistic or dust-like matter, points to an instant of creation or initial singularity. This is usually referred to as the big bang origin of the universe. The Hubble's discovery was in fact an observational verification of the expanding universe model. Later observations, such as the abundance of light elements in the universe, the cosmic microwave background radiation (CMBR), etc. were considered to endorse this picture. However,  this idea has found many ups and downs and many U-turns in the past eight decades and is an  example of the curious way scientific theories evolve. 

A  lacuna in modern cosmology is that the errors in the measurement  of the  present values of the expansion rates, related to the Hubble parameter $H_0$,  deceleration parameter $q_0$,  etc., of the universe are notoriously high. For example, Hubble's own measurement of $H_0$ is higher than the presently favored  values by an order of magnitude. Even the most optimistic measurements of this, quoted in recent works, have more than 5 percent error. Treatises on cosmology deal with the uncertainty in the measured value of $H_0$ by putting it as $H_0 =100 h $ km s$^{-1}$ Mpc$^{-1}$ and leaving $h$ in the expressions as some  undetermined parameter. The age of the universe has been another common stumbling block for the big bang theory. Many-a-time, it happens that the age of the universe in the big bang model is  smaller than the ages of several celestial objects when measured directly. An easy way to solve this age problem is to include a small but positive cosmological constant $\Lambda$ into the Einstein equation. The cosmological constant has a   repulsive effect  at large scales but at the same time it is assumed to have no other interactions with matter. (Energies with the latter property are sometimes termed as `dark energy', when they are assumed to be present in substantial amount at present.) By its inclusion one   obtains a larger age for the universe and thus can explain away the problem related to it.

In spite of all these difficulties,  many cosmologists believed  that the big bang model provided a reliable and tested accounting of the universe from about 0.01 sec after the Bang until today, some 15 billion years later \cite{kolbturner}. Thus in the 1980s, cosmologists were primarily occupied with some conceptual issues with the  big bang model,  which are referred to as the `cosmological problems'. The solution to some of these problems was found in the theory of inflation, an exponential expansion of the universe that might have occurred in its very early history, which can then lead to a flat universe. The energy which is supposed to have driven the inflationary expansion was a scalar field potential energy, which is akin to that of a cosmological constant, since it has an equation of state approximately of the form $p_{\Lambda}=-\rho_{\Lambda}$. But in the early 90s, there were no reasons to include  a dark energy in the present universe. The euphoria regarding the  validity of the big bang picture reached the pinnacle when the cosmic background explorer (COBE) satellite measured the fluctuations in the CMBR in 1992 to get its magnitude close to that predicted by the model. 

A recent development in theoretical cosmology is that the energy corresponding to the $\Lambda$-term or dark energy  has again come back to the present universe scenario, as it appeared in the literature several times in the past for various reasons.  Its present phase began with the measurement of the Hubble parameter  reported in \cite{freedpi,pierce},  that gave a   specific  and high range $h>0.7$. This, along with the flatness imposed by the theory of inflation, created a short spell of  an age problem,  which required a $\Lambda$-term for its solution. Then came the Type Ia Supernovae (SNe Ia) data \cite{per99,rie98}, which in turn required the presence of a $\Lambda$-term large enough to cause an accelerated expansion. The most recent release of SNe data \cite{riess1,perl1,riess04} prompts many cosmologists even to speculate that  some extremely unphysical energy densities, such as those with $w<-1$ in the equation of state  $p=w\rho$ are required to explain the data. 

Thus unlike in the past, it seems that the suspected dark energy in the universe cannot  be regarded as a  cosmological constant of general relativistic origin or as the vacuum energy encountered in quantum field theories. Our knowledge of the physical world then remains deficient since no tested theory involves such a dark energy. But in this talk I wish to caution that attempting to analyze the new SNe data by assuming exotic energy densities with strange equations of state itself is very misleading and argue that the reasonable remaining  option is to analyze them  cosmographically, without making any specific assumptions on the energy densities in the universe. Such an analysis of SNe data in \cite{mvj1} assumes only homogeneity and isotropy for the universe; i.e., the universe is assumed to have a Robertson-Walker (RW) metric. The  scale factor of the universe can most naturally be expanded into a  Taylor series in time about the present epoch and we attempt to find its coefficients from observation, as it was done to evaluate the Hubble parameter and deceleration parameter in the original cosmographic approach. Terms up to fifth-order were kept in the above series and it was assumed that they make a good approximation. The likelihoods for the various expansion rates obtained in our calculation are very broadly peaked  and these indicate that there are a variety of choices for the energy densities in the present  epoch. However, also this model-independent  analysis confirmed that the universe is accelerating at present.

An important question I try to answer in this talk is whether the data really endorse that   the universe was decelerating  in the past \cite{mvj2}. The answer has serious implications for modern theoretical cosmology for it is almost entirely built on the firm belief that the universe was decelerating in the past. Theories of the early universe, such as those of nucleosynthesis, CMBR, formation of structures, etc., were developed before the 1990s, under the assumption that the universe was always decelerating. Now with the arrival of the new SNe data, there appears to be a consensus that the universe is accelerating at present, and that this may be due to substantial amount of dark energy. But it should be noted that an accelerating universe was not in anyone's wildest imagination, at any time before these SNe observations were made. Obviously, this is sufficient reason to suspect  the validity of the entire early universe theories. When the theoretical and observational status of the present universe has suffered such a setback, it would be naive to believe that the theories and observations regarding the early universe remain intact. The attempts of some cosmologists to deduce from the SNe data the redshift at which universe changed from deceleration to acceleration  can only be viewed prima facie as a damage control measure. My attempt in this paper is to inspect this claim, whether the  the latest SNe data give any indication that the universe changed from deceleration to acceleration at any foreseeable past!

To answer our question,   the  Bayesian method of model comparison is adopted in which a model with some decelerating phase in the past is compared with another one having no such phase. In both cases, we assume the present universe to be accelerating.  It is claimed that this method of comparison is more robust than other investigations seeking evidence for a decelerating past since  both the Bayesian and cosmographic approaches mentioned above are used here. As suggested earlier,   the scale factor is expanded into a fifth-order polynomial  in time (fifth-order is required for sufficient accuracy) and then the  combinations of various coefficients in this expansion  were separated into those which correspond only to acceleration during the entire period and those which had at least some decelerating phase in this period. Considering these as rival models, the Bayes factor is calculated. The results show that if at all there is evidence, it is in favor of an always accelerating model; at best one can say that the SNe data are not  capable of discriminating   the ``always accelerating" and ``decelerating in the past" cosmological models.

The talk is organized as follows. In Section 2, we state the standard methods to obtain the apparent magnitude-redshift relation of objects and the best fit parameter values in a model. Section 3 gives an overview of the Bayesian model comparison technique. It is in Section 4 that we present the Bayesian, model-independent analysis of our problem. The last section comprises a discussion of our results.

\section{ THE APPARENT MAGNITUDE-REDSHIFT RELATION }

The conventional method of analyzing the apparent magnitude-redshift data relies on first solving the Einstein equations in cosmology for $a(t)$, the scale factor appearing in the line element 

\begin{equation}
ds^{2} = dt^{2} - a^{2}(t) \left[ \frac {dr^{2}}{1-kr^{2}} +
r^{2} (d\theta ^{2} + \sin ^{2}\theta\; d\phi ^{2})\right] .
\label{eq:rwle} 
\end{equation}
These equations are

\begin{equation}
\frac {\dot {a}^{2}}{a^{2}} +\frac {k}{a^{2}} = \frac {8\pi
G}{3} \rho, \label{eq:t-t}
\end{equation}
and

\begin{equation}
2 \frac {\ddot{a}}{a} + \frac {\dot{a}^{2}}{a^{2}} +\frac {k}{a^{2}} =
 -8\pi G p. \label{eq:s-s}
\end{equation}
Here $p$ is the pressure and $\rho$ is the density. We also need an equation of state of the form $p=w\rho $ to solve the above. ($w$ is called the equation of state parameter.)
 For each model,  the solution of these equations helps us to predict the $m$-$z$ relation as follows. Put $ds=0$ in Eq. (\ref{eq:rwle}) for light coming from an SN, which is at position $r_1$ at time $t_1$, and reaches us at $r=0$ at the present time $t_0$ and then integrate

\begin{equation}
\int_{t_1}^{t_0} \frac{dt}{a(t)} = \int_{r_1}^{0} \frac{dr}{(1-kr^2)^{1/2}}. \label{eq:int}
\end{equation}
This helps to find $r_1$ or $r_1a_0$ using the solution $a(t)$ of Eqs. (\ref{eq:t-t}) and (\ref{eq:s-s}) obtained for the  model under consideration. Then we  compute the luminosity distance $D=r_1a_0(1+z)$ and from it the apparent magnitude

$$
m=5 \log \left(\frac{D}{1 \hbox {Mpc}}\right)+25+M.
$$
Here we should make use  of the relation  
$$
1+z= \frac{a(t_0)}{a(t_1)},
$$
to obtain $t_1$ in terms of $z$. It may be noted that 
 $D$ and hence $m$ can be functions of  $M$, $a_0$, $H_0$, $k$, and other parameters in $a(t)$ such as $\Omega_m \equiv \rho_{m}/\rho_{c}$, $\Omega_{\Lambda} \equiv \rho_{\Lambda}/\rho_{c}$, etc., depending upon the model chosen.

The evaluation of the best fit values of the parameters $\Omega_{m} $, $\Omega_{\Lambda} $, etc., in a model is usually based on  the minimum of the $\chi^2$-statistic

\begin{equation}\label{eq:chi2}
\chi^2 = \Sigma _k \left[ \frac{\hat{m}_k -m_k(M, H_0, \Omega_{m} , \Omega_{\Lambda} , ..)}{\sigma_k}\right]^2,
 \end{equation}
where $\hat{m}_k$ denotes the measured value of the apparent magnitude of the $k^{th}$ SN, $m_k(M, H_0, \Omega_{m} , \Omega_{\Lambda} , ..)$  its expected values (from theory) and $\sigma _k$  the uncertainties in the measurement of the observables.

\section{BAYESIAN MODEL COMPARISON}

The Bayes's probability  theorem tells us how to adjust our plausibility assumptions regarding a hypothesis when our state of knowledge changes through the acquisition of new data.  Classical mathematicians such as Bernoulli, Bayes, Laplace and Gauss have found Bayes theorem useful in problems such as those in astronomy, thanks to its ability to learn. For example, Laplace has used this theory to estimate the masses of planets and to quantify the uncertainty of the masses due to observational errors. Later, because of the difficulties with assigning prior probabilities (which were mistakenly considered to be purely subjective expressions of a person's opinions about hypotheses), Bayes's probability theory has gone out of favor in physical sciences  and was replaced by the more apparently objective `frequentist' probability approach. But the realization that the frequentist definition of probability is as subjective  as the Bayesian has called forth a re-examination of this controversy. When compared to the improper application of frequentist probability theory, the Bayesian approach is most powerful in   problems such as those in astronomy, where the process of learning and also the quantification and readjustment of plausibility by the scientific community are very important.

In the Bayesian model comparison approach \cite{mvjvn}, we consider  two models   (say, $M_i$ and $M_j$) as rival hypotheses and try to compare them by evaluating odds ratios between the posterior (i.e., after analyzing the data) probabilities $p(M\mid D,I)$ for these models, given the data $D$ and also assuming the truth of some background information $I$. For this, the Bayes's theorem is made use of.
This theorem states that the posterior  probability  for a hypothesis $H_i$ given  $D$ and  $I$ is

\begin{equation} \label{eq:prob}
p(H_i|D,I)= \frac {p(H_i|I)p(D|H_i,I)}{p(D|I)}.
\end{equation}
Here $ p(H_i|I) $ is called the prior  probability  for $H_i$, given the truth of $I$ alone. $p(D|H_i,I)$ is the probability for obtaining the data $D$ if the hypothesis $H_i$ and $I$ were true and is called the likelihood for the hypothesis. The factor in the denominator serves the purpose of normalization.

Considering the truths of models as our hypotheses, the Bayes' theorem helps us to write the odds ratio between  $M_i$ and $M_j$, which is the ratio between the corresponding posteriors, as

\begin{equation}\label{eq:odds}
\frac{p(M_i|D,I) }{p(M_j|D,I)}= \frac {p(M_i|I)p(D|M_i,I)} {p(M_j|I)p(D|M_j,I)} \equiv O_{ij}.
\end{equation}

If the information $I$ does not prefer one model over the other, the prior probabilities get canceled out and the odds ratio reduces to $B_{ij}$, which is called the Bayes factor.

\begin{equation}\label{eq:bayes}
 O_{ij}= \frac {p(D|M_i,I)} {p(D|M_j,I)} \equiv B_{ij}.
\end{equation}
For parameterized models, the probability for the data given the truth of the model, also called the likelihood for the model, can be written as

$$
p(D|M_i,I)\equiv {\cal L}(M_i)=\int d\alpha \int d\beta ... p(\alpha , \beta , ...|M_i) {\cal L}_i (\alpha , \beta , ...).
$$
Here, $p(\alpha , \beta , ...|M_i)$ is the prior probability for  $\alpha , \beta , ..$ and the likelihood for the parameters in the model $M_i$ is given by

\begin{equation}\label{eq:likelipar}
{\cal L}_i (\alpha , \beta , ...)=\exp \left[-\chi _i^2(\alpha , \beta , ..)/2 \right].
\end{equation}

 The commonly accepted interpretation of the Bayes factor is as follows: If $1<B_{ij}<3$, there is evidence against model $M_j$, but it is not worth more than a bare mention. If $3<B_{ij}<20$, this evidence is positive. If $20<B_{ij}<150$, it is strong and if $B_{ij} >150$, the evidence is very strong.

As familiar examples of cosmological models, we consider the following for  illustrating the Bayesian model comparison technique.

(1) Friedman-Lamaitre-Robertson-Walker (FLRW) model

In this model, the universe contains two components for the total energy density $\rho$; the first corresponds to ordinary matter and the second to a cosmological constant $\Lambda$. i.e., 

$$
\rho = \rho_m + \rho_{\Lambda}, \qquad p=p_m + p_{\Lambda},
$$

$$
\Lambda =constant, \qquad \rho_{\Lambda} = \frac{\Lambda}{8\pi G},
$$

$$
p_m=0, \qquad p_{\Lambda}= -  \rho_{\Lambda}.
$$

(2) FLRW model with inflation (Flat universe)

The same as the above with the additional condition $ \Omega \equiv \Omega_m + \Omega_{\Lambda} =1$, which reduces the effective number of parameters by unity.

Bayes factors between FLRW models  with and without inflation (Models $M_1$ and $M_2$ respectively) obtained in \cite{mvjvn} are

$B_{12} = 1.6$ using the  data in \cite{per99} (Fit C) 

$B_{12} = 1.1$ using the  `All SCP SNe' data in \cite{perl1} 

This example shows that the inflationary model is slightly more favored than the general FLRW model, but the evidence is not worth more than a bare mention. When the Bayes factors are close to unity like this, it is highly objectionable to state that one or the other model is `ruled out' by the data. It may also be reminded that the price for overlooking this fact would be frequent and unavoidable U-turns.

\section{THE MODEL-INDEPENDENT APPROACH TO THE PROBLEM}

In this section, we examine the claim that the universe changed from deceleration to acceleration at some time in the past, by adopting the Bayesian, model-independent  approach.

As stated in the introduction, the basic assumption in the model-independent or cosmographic approach is that the the scale factor of the universe can be approximated by the first six terms in a Taylor expansion \cite{mvj1, mvj2}. i.e.,

\begin{eqnarray}
a(t_0+T)&=&a_0\left[
1+H_0T-\frac{q_0H_0^{2}}{2!}T^2+
\frac{r_0H_0^3}{3!}T^3-  \frac{s_0H_0^4}{4!}T^4 +\frac{u_0H_0^5}{5!}T^5
\right]  \nonumber \\
&\equiv& a_0\left[1+a_{(1)}T+a_{(2)}T^2+a_{(3)}T^3+a_{(4)}T^4+a_{(5)}T^5\right] 
\label{eq:a1}.
\end{eqnarray}

We shall use this expression in the integral in Eq. (\ref{eq:int}). Parameters in this case are $a_0$, $h$, $q_0$,  $r_0$, $s_0$, $u_0$, $M$, and $k$. We assume flat priors for  the parameters in the ranges $-19.6<M<-19.1$ magnitudes, $0.6<h<0.8$, $-2<q_0<0$, $-15<r_0<15$, $-65<s_0<65$, and $-150<u_0<150$, $ 8000>a_0>3000 $ Mpc, and $k=\pm 1$. These ranges are as obtained in \cite{mvj1}.

The two hypotheses we want to compare are (1) the universe is always accelerating from time $t_0+T_{p}$  to the present epoch $t_0$ (model $M_1$) and (2) there is at least one decelerating epoch for the universe during this period (model $M_2$). For any particular combination of parameter values,  a  sure test for the occurrence of deceleration  during $T_{p}<T<0$ is to plot

\begin{equation}\label{eq:addot}
\ddot{a}(t_0 +T)=a_0[2a_{(2)}+6a_{(3)}T+12a_{(4)}T^2+20a_{(5)}T^3]
\end{equation}
for this interval and to see whether it becomes negative at any time during the period.

The values of the Bayes factor obtained while using the 54 ``All SCP" SNe (data $D_1$) in  \cite{perl1}  (as reproduced in \cite{mvj1}) and the 157 ``gold" data points (data $D_2$) in \cite{riess04}  are given in Table 1. The results show that except in one case, the Bayes factors favor the `always accelerating' Model 1. But in all cases, the evidences are  weak.

\begin{table}
\begin{tabular}{ccc}
\hline
$T_{p}$ s &  $B_{12}$ using $D_1$ &  $B_{12}$ using  $D_2$ \\ \hline
$-1\times 10^{17}$ &3.3 & 1.1 \\
$-2\times 10^{17}$ &1.5 & 0.6 \\
$-3\times 10^{17}$ &2  & 1.1 \\
$-4\times 10^{17}$ &2.4 & 1.6 \\
$-5\times 10^{17}$ &2.8  & 2.1\\ \hline
\end{tabular}
\caption{Bayes factors between  the ``always accelerating" ($M_1$) and ``decelerating in the past" ($M_2$) cosmological models.}
\label{tab:a}
\end{table}

\section{DISCUSSION}

We see that there is hardly any evidence in support of a decelerating past for the universe in the past 15 Gyrs, except in the case with $T_p=-2\times 10^{17}$ s and data $D_2$. These results are obtained using a combined Bayesian, model-independent approach. However,  as per the interpretation of the Bayes factor given in Sec. 3, the evidences against a decelerating past for the universe are in general weak.  Thus it may safely be concluded that these data alone cannot discriminate the two hypotheses we compared. In summary, we state that even the most precise cosmological data, which is that of the SNe Ia, cannot provide any clue as to whether the universe changed from an earlier decelerating phase to the present accelerating one at some time in the past. 

Thus we are still at the starting point in cosmology, where the expansion of the universe is the only authentic piece of information, validated by known physics. 

\begin{theacknowledgments}
  
I am thankful to the Organizing Committee of the 1$^{st}$ Crisis in Cosmology Conference for travel and hospitality. I also wish to thank the University Grants Commission, New Delhi for a research grant.

\end{theacknowledgments}


\end{document}